%% file: main.tex
\documentclass[conference]{IEEEtran}
\usepackage{cite}
\usepackage{float}
\floatstyle{plaintop}
\restylefloat{table}
\usepackage{listings}
\usepackage{caption} 
\usepackage{adjustbox}
\usepackage{graphicx}
\usepackage{array}
\usepackage{amsmath}
\usepackage{xcolor}
\graphicspath{{figures}{figures/images}{figures/screenshots}}
\DeclareGraphicsExtensions{.pdf,.jpeg,.png}
\usepackage{amsmath}
\usepackage{tabularx}
\usepackage{array}
\usepackage{booktabs}
\ifCLASSOPTIONcompsoc
\usepackage[caption=false,font=normalsize,labelfont=sf,textfont=sf]{subfig}
\else
\usepackage[caption=false,font=footnotesize]{subfig}
\usepackage{latexsym}
\usepackage{rotating}
\begin{document}

	
	\title{Time-Shared Execution of Realtime Computer Vision
          Pipelines by Dynamic Partial Reconfiguration}

%	\title{Executing Multiple Realtime Vision
%          Pipelines by Time-Sharing using Dynamic Partial Reconfiguration}
	
	%	\title{Sharing Resources in Dynamically Reconfigurable Field Programmable Gate Arrays Systems to Reduce Fragmentation across Reconfigurable Regions}
	
	% use a multiple column floorplan for up to three different

		\author{\IEEEauthorblockN{Marie Nguyen}
			\IEEEauthorblockA{Carnegie Mellon University\\
				Pittsburgh, Pennsylvania \\}
			\and
			\IEEEauthorblockN{James C. Hoe}
			\IEEEauthorblockA{Carnegie Mellon University\\
				Pittsburgh, Pennsylvania}
	%%		%\and
	%%		%\IEEEauthorblockN{James Kirk\\ and Montgomery Scott}
	%%		%\IEEEauthorblockA{Starfleet Academy\\
	%%		%San Francisco, California 96678--2391}\\
	%%		%Fax: (888) 555--1212}}
		}

	
	%Telephone: (800) 555--1212, Fax: (888) 555--1212}

\def\mypar/#1{{\noindent \bf{#1.}}}
\def\james/#1{{\color{blue} {\em (jch: #1)}}}
\def\marie/#1{{\color{purple} {\em (marie: #1)}}}
	
\def\RP/#1{${\sf{RP}}_{\sf{#1}}$}
\def\MM/#1{${\sf{M}}_{\sf{#1}}$}
\def\PP/#1{${\sf{P}}_{{#1}}$}

\def\Tround/{$T_{{\sf{round}}}$}

\def\TslicePI/#1#2{$T_{{\sf{slice,{#1}}}_#2}$}
\def\TsliceP/#1{$T_{{\sf{slice,{#1}}}}$}
\def\TconfigPI/#1#2{$T_{{\sf{config,{#1}}}_#2}$}
\def\TconfigP/#1{$T_{{\sf{config,{#1}}}}$}

\def\TfillP/#1{$T_{{\sf{fill,{#1}}}}$}
\def\TfillPMath/#1{T_{{\sf{fill,{#1}}}}}
\def\TfillPI/#1#2{$T_{{\sf{fill,{#1}}}_#2}$}
\def\TfillPIMath/#1#2{T_{{\sf{fill,{#1}}}_#2}}
\def\TfillNo/{$T_{{\sf{fill}}}$}

\def\TframeP/#1{$T_{{\sf{frame,{#1}}}}$}
\def\TframePMath/#1{T_{{\sf{frame,{#1}}}}}
\def\TframePI/#1#2{$T_{{\sf{frame,{#1}}}_#2}$}
\def\TframePIMath/#1#2{T_{{\sf{frame,{#1}}}_#2}}
\def\TframeNo/{$T_{{\sf{frame}}}$}

\def\TfillrawP/#1{$T_{{\sf{fill{\text{-}}solo,{#1}}}}$}
\def\TfillrawPMath/#1{T_{{\sf{fill{\text{-}}solo,{#1}}}}}
\def\TfillrawPI/#1#2{$T_{{\sf{fill{\text{-}}solo,{#1}}}_#2}$}
\def\TfillrawPIMath/#1#2{T_{{\sf{fill{\text{-}}solo,{#1}}}_#2}}
\def\TfillrawNo/{$T_{{\sf{fill{\text{-}}solo}}}$}

\def\TframerawP/#1{$T_{{\sf{frame{\text{-}}solo,{#1}}}}$}
\def\TframerawPMath/#1{T_{{\sf{frame{\text{-}}solo,{#1}}}}}
\def\TframerawPI/#1#2{$T_{{\sf{frame{\text{-}}solo,{#1}}}_#2}$}
\def\TframerawPIMath/#1#2{T_{{\sf{frame{\text{-}}solo,{#1}}}_#2}}
\def\TframerawNo/{$T_{{\sf{frame{\text{-}}solo}}}$}

	\maketitle
	
	
	\input{0_abstract}


	% \fi
	\IEEEpeerreviewmaketitle
	
\input{1_intro}

\input{2_background}

\input{3_timeshare}

\input{4_techniques}

\input{5_prototype}

\input{6b_results}

\input{7_discussion}
\input{8_acknowledgments}
%\input{8_conclusion}

%\balance
	
%\nocite{Chung:2011:CIM:1950413.1950435}
	
	\bibliographystyle{ieeetr}
	\bibliography{references}
	
	\newpage
	%	\input{unused}
	%	\input{problem}
	%	\input{implement}
	
	
	
	%\bibliographystyle{IEEEtran}
	
	%\input{intro_}
	%\input{motivation_}
	%\input{solution}	
	%	\input{implementation}
	%\input{results}
	%	\input{related}
	%\input{conclude}	
	
\end{document}

%% file: 0_abstract.tex
\begin{abstract}
	
	This paper presents an FPGA runtime framework that demonstrates the feasibility of using
	dynamic partial reconfiguration (DPR) for time-sharing an FPGA by multiple realtime computer vision pipelines. The presented time-sharing runtime framework manages an FPGA fabric that can be round-robin time-shared by
	different pipelines at the time scale of individual frames. In this new use-case, the challenge is to achieve useful performance despite high reconfiguration time. The
	paper describes the basic runtime support as well as four optimizations
	necessary to achieve realtime performance given the limitations of DPR on
	today's FPGAs. The paper provides a characterization of a working runtime framework prototype on a Xilinx ZC706 development board. The paper
	also reports the performance of realtime computer vision pipelines when time-shared. 
	
%	\marie/{@james: I changed the following sections:
%	 - background (I added references in the DPR subsection)
%	 - prototype (removed the synthetic benchmark results)
%	 - performance evaluation (show flexibility of framework by time sharing many more pipelines than can fit at once)
%	 - conclusion
%	 I added another section (discussion).}
	\noindent
	{\bf{keywords:}} partial reconfiguration, realtime time-sharing, streaming, vision.
\end{abstract}

%% file: 1_intro.tex
\section{Introduction}
\label{sec:intro}

\mypar/{Motivation} %
%FPGAs have been applied extensively in vision
%processing on realtime video streams~\cite{5159427,
%	Hegarty:2014:DCH:2601097.2601174,
%	Hegarty:2016:RFM:2897824.2925892}. 
FPGAs have increasingly been deployed in compute settings. However, past examples have not fully
exploited the dynamic programmability of FPGAs. Typically, once a
design is loaded, the FPGA acts as an ASIC with a fixed set of
functionalities that are statically mapped on the FPGA for the duration of the deployment. 
%Also like an
%ASIC, all included functionalities consume logic resources and power whether or
%not a functionality is in active use.

Modern computer vision applications have become interactive,
requiring systems to adapt dynamically in
functionality and/or in performance to user and environment inputs. For instance, advanced driver-assistance systems (ADAS) need to change the behavior of the car in realtime based on the driver and on sensor inputs. 
These interactive applications present an opportunity to leverage FPGAs dynamic programmability potential. The main challenge when implementing interactive realtime systems is that the sequence and combination of applications requested at runtime are not known at design time. This dynamic adaptation requirement leads to a very large number of potential FPGA states.
%Implementing such a system in a traditional static design is inflexible, expensive (in terms of device and power cost), and may even be impossible.  
Mapping all possible application combinations on an FPGA using a traditional static design flow is inflexible, expensive and may be impossible given the area or power budget alloted. Also, statically mapping all possible applications combinations is wasteful since only a subset of applications needs to be active at a time.

%Modern vision applications have become interactive,
%requiring vision systems to adapt dynamically in
%functionality and/or in performance to user and environment inputs.
Dynamic partial reconfiguration (DPR) has been used successfully to
provide on-the-fly adaptability by allowing an FPGA to be repurposed
with new functionalities with minimal operational disruption~\cite{1303106}. In the context of repurposing for interactivity or adaptability,
the interval between reconfiguration is in the minutes to hours range
with tolerance for missed frames during reconfiguration.  For these
uses, the FPGA still acts like an ASIC for extensive periods in
between reconfigurations. 

The work in this paper aims to {\em{apply DPR to time-share the FPGA fabric by multiple realtime computer vision pipelines at
		the time scale of individual frames.}} 
With time-sharing, the FPGA can support more concurrent realtime
functionalities than what could statically fit on the FPGA. In this use-case, every frame must be processed by every pipeline.

% In this paper, we present a new use case for DPR where performance 
%Despite the high DPR reconfiguration time,  this paper presents a new case of DPR for vision processing where performance requirements are within the millisecond range.  

\begin{figure}[t]
	\centering 
	\includegraphics[scale=0.5]{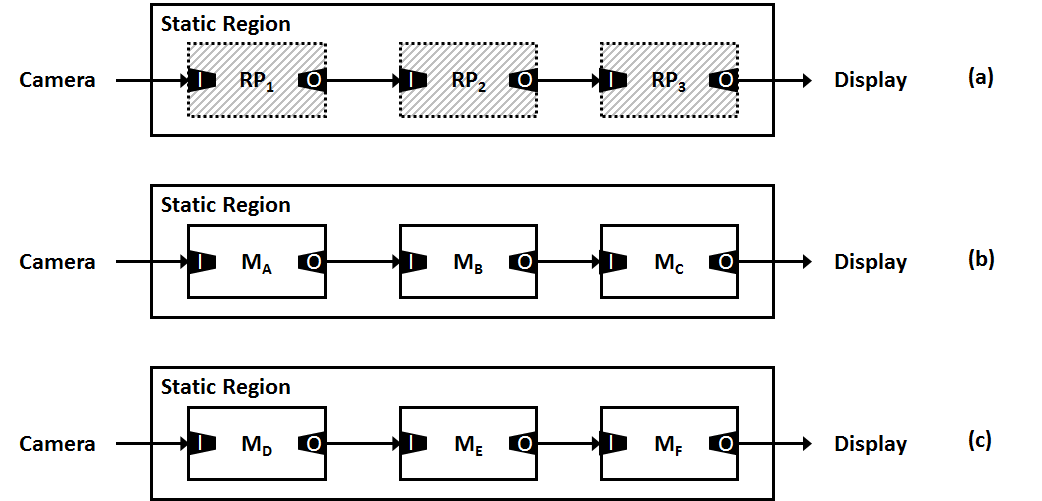}
	\caption{Conceptual sketch of a repurposable DPR framework for computer vision 
		pipelines. (a): the fabric is organized into a static
		region and three reconfigurable partitions
		(\RP/{1} $\sim$ \RP/{3}). (b)\&(c): Different computer vision pipelines can be
		executed over time by reloading the reconfigurable partitions with
		modules from a library.}
	\label{fig:dpr_example}
\end{figure}

\mypar/{Prior Work: DPR for Repurposing} DPR allows a region of the
FPGA fabric to be reconfigured without disrupting the
operation of the remainder of the fabric~\cite{xilinx}. As an
illustrative example, Figure~\ref{fig:dpr_example}.a depicts an FPGA
fabric organized into a static region and three reconfigurable
partitions (RPs). The static region provides infrastructure logic
to connect the camera to the first RP (\RP/{1}) and
the last RP (\RP/{3}) to display. The
infrastructural logic further connects the three RPs by streaming
connections in a linear topology. The RPs can be reconfigured with
pre-compiled modules (\MM/{A} $\sim$
\MM/{F}) from a library to serve as the stages of a computer vision pipeline. For example,
Figure~\ref{fig:dpr_example}.b shows the RPs configured as the
pipeline: {\sf{camera}} $\rightarrow$ \MM/{A} $\rightarrow$ \MM/{B}
$\rightarrow$ \MM/{C} $\rightarrow$ {\sf{display}}.  Alternatively,
Figure~\ref{fig:dpr_example}.c shows the RPs repurposed as a different
pipeline: {\sf{camera}} $\rightarrow$ \MM/{D} $\rightarrow$ \MM/{E}
$\rightarrow$ \MM/{F} $\rightarrow$ {\sf{display}}.  

In practice, a modern large FPGA could have 
more than three RPs, and a framework could support more elaborate
streaming connection topology between RPs. We have created a repurposable DPR framework as sketched above in
prior work. Others have also shown that this kind of repurposable DPR framework is readily
realizable using standard DPR support in commercial
tools and FPGAs~\cite{Majer:2007:ESM:1265130.1265134, autovision}. 

\begin{figure}[t]
	\centering
	\includegraphics[scale=0.65]{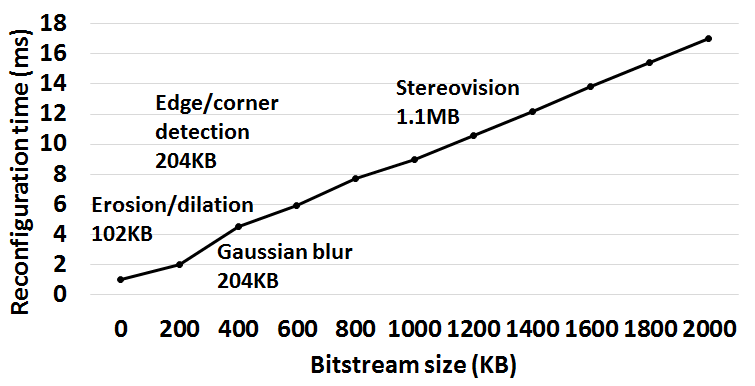}
	\caption{The measured reconfiguration time as a function of the RP size
		(represented by the size of its bitstream). Sizes of common computer vision
		modules are marked as reference points.}
	\label{fig:raw_overhead}
\end{figure}

\mypar/{This Paper: DPR for Realtime Time-Sharing}  In this work, we want to
apply DPR multiple times in the time scale of a single camera frame to
support time-shared round-robin execution of multiple realtime computer vision
pipelines.
For example, to time-share the two pipelines in
Figure~\ref{fig:dpr_example}.b and~\ref{fig:dpr_example}.c, we divide
the time quantum of one camera frame 
into two timeslices for both pipelines.  
The time scale for time-sharing in computer vision applications---just 16.7~milliseconds per frame at 60 frames-per-second (fps)---is just within reach of today's DPR support.
To maintain realtime processing, both
the reconfiguration time and the time for processing one camera frame have
to fit within a pipeline timeslice. Each camera 
frame is buffered for processing by the two pipelines
during their respective timeslice. The output frames from the two
pipelines could be merged for display (e.g., split-screen) or buffered
separately in DRAM for downstream consumption. By synchronizing
time-sharing with frame boundaries, there is no context (frames) to save/restore when switching between pipelines.

The challenge when time-sharing comes from the restrictive DPR speed in today's FPGAs.  On the Xilinx XCZ7045 FPGA, the time to reconfigure one RP is proportional to its size,
typically a few to 10s of milliseconds (see
Figure~\ref{fig:raw_overhead}).  Moreover, only one RP can be
reconfigured at a time so the multiple RPs of a pipeline are
reconfigured one after the other.  Consequently, the
reconfiguration time alone can easily overwhelm the alloted time slice at any non-trivial frame rates.

\mypar/{Contributions} In this work, we show that it is feasible to use DPR for time-sharing realtime computer vision pipelines with performance requirements in the tens of milliseconds range. We develop four optimizations to hide, amortize or eliminate reconfiguration time, and are necessary to achieve usable time-sharing performance for computer vision pipelines. We have created a runtime framework for computer vision streaming processing that implements these optimizations: 
\begin{itemize}
	\item overlapping stage reconfiguration and processing within a pipeline timeslice to hide reconfiguration time
	\item round-robin scheduling at an enlarged granularity of multi-frame bundles to amortize reconfiguration cost over the processing time of multiple frames
	\item a flexibly configurable streaming interconnect infrastructure to reduce the number of partitions that must be reconfigured when pipelines share common stages
	\item downsampling video stream from camera (with lower effective frame rate) in a fashion transparent to the computer vision modules instantiated into the RPs
\end{itemize}
%First, the runtime framework supports overlapping stage reconfiguration and processing within a pipeline timeslice to hide reconfiguration time. Second, the runtime framework offers the option of round-robin scheduling at
%an enlarged granularity of multi-frame bundles to amortize reconfiguration cost over the processing time of multiple
%frames. Third, the runtime framework provides a flexibly configurable
%streaming interconnect infrastructure to reduce the number of
%partitions that must be reconfigured when pipelines share common stages. %Fourth, the runtime framework can drive
%the pipelines with a down-sampled video stream from camera (with lower effective
%frame rate) in a fashion transparent to the vision modules instantiated into the RPs.
%We have created a working runtime framework for vision streaming processing to demonstrate that the feasability of time-shared execution.
Our final results show that we can time-share an FPGA between streaming computer vision pipelines, and achieve useful frame rates (30+~fps) for each time-shared pipeline.

\mypar/{Paper Outline} Following this introduction,
Section~\ref{sec:background} provides background and a survey of
related work. Section~\ref{sec:timeshare} presents the basic
design and operation of our time-sharing runtime framework.
Section~\ref{sec:techniques} next presents the techniques to reduce
the impact of reconfiguration time in time-sharing. Section~\ref{sec:solution}
describes a working realization of the presented runtime framework on a Xilinx
ZC706 development board. Section~\ref{sec:results} presents the
performance evaluation when time-sharing streaming computer vision pipelines. Lastly, Section~\ref{sec:discussion} offers our
conclusions.

% and suggestions for future work and future changes.

%% file: 2_background.tex
\section{Background}
\label{sec:background}

\mypar/{Dynamic Partial Reconfiguration (DPR)} Section~\ref{sec:intro}
briefly introduced the notions of a static region and reconfigurable
partitions (RPs) in a DPR example
(Figure~\ref{fig:dpr_example}). In Xilinx's environment, at design
time, the RPs appear as black-box submodules with declared
input/output ports but unspecified internals.  At implementation time,
the RPs' bounding box and port locations are fixed by floorplanning.
Separately, different modules that (1)~have matching
interface ports and (2)~can fit within the logic resources of a RP can
be separately placed-and-routed as variants to be loaded into the RP
at runtime. In our work, we control reconfiguration from the embedded ARM core.
The bitstreams of the modules are stored and loaded at runtime from DRAM to reconfigure 
the RPs through Xilinx's PCAP interface~\cite{xilinx}.

DPR technology has been supported for over a decade and has been
used in many prior works (e.g., \cite{Koch:2011:FHP:1950413.1950427,
	6128547, 6339136}). In these works, DPR has been exploited for saving
area by time multiplexing application phases~\cite{Arram:2015:RRA:2684746.2689066}, for customizing data paths to
improve performance~\cite{Niu:2015:EOC:2684746.2689076}, or for
virtualizing FPGA resources in the cloud~\cite{6861604}. DPR has been used for streaming vision processing~\cite{Majer:2007:ESM:1265130.1265134, autovision} at a coarse time-scale, where the time between reconfigurations is within seconds to minutes range, to repurpose the FPGA for different functionalities. In \cite{4630015}, the authors identified the issues of asynchronous module execution and frame skipping when applying DPR for repurposing in a vision streaming context. 

%This paper focuses on the mechanisms and techniques to support realtime time-sharing for streaming vision pipelines. 
%(when the time between reconfigurations is within milliseconds range) while achieving useful performance in a streaming vision context.

\mypar/{Streaming Vision Pipeline} We use the simple streaming vision pipelines depicted in Figure~\ref{fig:dpr_example} to explain the operation of standard streaming vision pipelines and the operation of time-sharing. We will discuss the operation of more complex pipelines later in the paper (Figure~\ref{fig:pipe_eval}).
%Later in the paper will show more
%complicated pipelines (Figure~\ref{fig:pipe_eval}) that can be handled
%by our implementation.  For now, we continue to base the presentation
%on a simple conceptual model of streaming vision pipelines like in
%Figure~\ref{fig:dpr_example}.  
We assume the streaming vision pipeline is driven by a camera and outputs to a display, and that pixels are continuously streamed into the pipeline. The camera streams pixels into the first stage of the pipeline at a steady rate.
\TframeP/{camera} is the time between the first pixel and last pixel of
a frame produced by a camera; the frame rate is
$\frac{1}{\TframePMath/{camera}}$. In a simple pipeline, all pipeline stages consume and
produce pixels at the same steady rate as the camera, logically
computing an output frame from each input frame. The stages may need
to buffer multiple lines of the frame but never a complete frame. Due
to buffering, there is a delay between when the first pixel of a frame
enters a stage (or a pipeline) and when the first pixel of the
same frame exits a stage (or a pipeline); this time is
\TfillP/{stage}. After the first pixel exits, the last pixel exits
\TframeNo/ later. In steady-state with continuous streaming inputs, a complete frame would exit every
\TframeNo/. 

For example, considering a full-HD camera that outputs frames with 1920-by-1080 pixels at 60 frames-per-second (fps), \TframeP/{camera}=16.7 milliseconds. When
using a 16-bit wide streaming connection, the pipeline needs to operate at a minimum frequency of 148.5~MHz. If a stage requires buffering of 10 lines of the frame, \TfillNo/ of that stage will be at least 0.13 milliseconds
($10~{\text{lines}} \times \frac{1920~\text{pixels}}{\text{line}} \times
\frac{cyc}{16~{\text{pixels}}} \times \frac{1}{148.5~\text{MHz}}$). (The size of a full-HD frame in YUYV422 format (2 bytes/pixel) is 4~MB.) 

Under basic operation, any given stage just needs to keep up with the
pixel rate from the camera. However, a stage running by itself could be
clocked faster resulting in shorter \TframerawP/{stage} and
\TfillrawP/{stage}. For example, this is applicable when the streaming input and output of a stage are
sourced from and sinked into DRAM instead of camera and display.

%% file: 3_timeshare.tex
\section{Basic Round-Robin Time-Sharing}
\label{sec:timeshare}

This section describes the operation of a basic time-sharing
system and its performance model. 

\begin{figure}[t]
	\centering
	\includegraphics[width=0.5\textwidth]{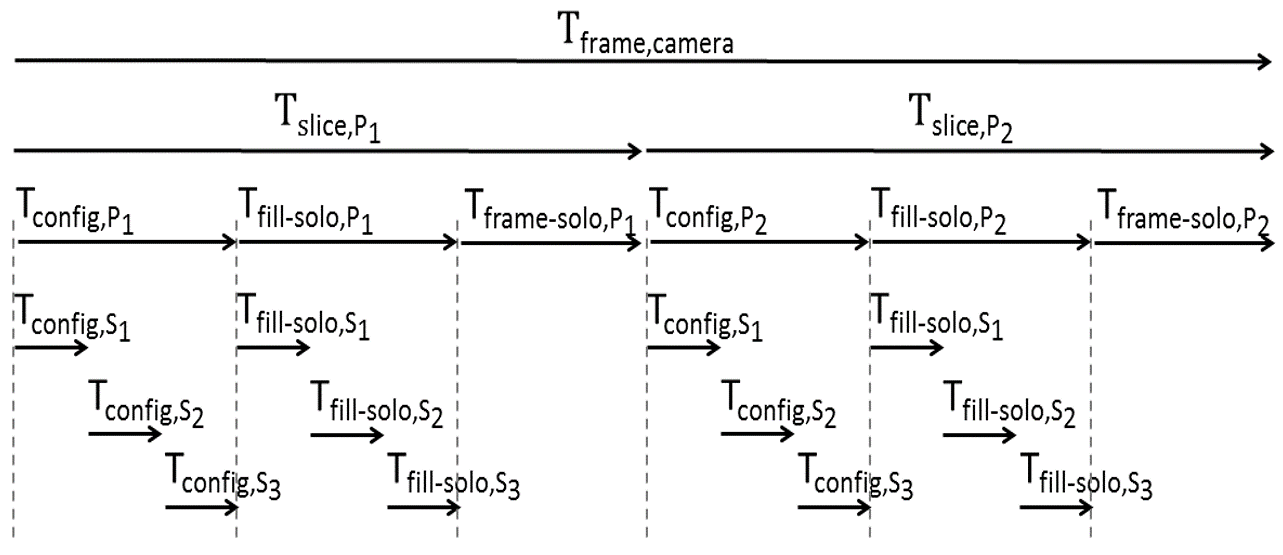}
	\caption{Time-sharing by two three-stage pipelines.  A pipeline starts processing only 
		after all stages have been configured.}
	%	\caption{This time line shows the time interval between two
	%	consecutive start of frames ($T_\text{frame,camera}$). This
	%	time interval is split into two pipeline timeslices
	%	$timeslice_{\sf{P{_{1}}}}$ and
	%	$timeslice_{\sf{P{_{2}}}}$. During each timeslice, a pipeline
	%	(\PP/{1} or \PP/{2}) is reconfigured (by reconfiguring three
	%	reconfigurable partitions \RP/{1}, \RP/{2} and \RP/{3}
	%	sequentially) and processes one frame. }
	\label{fig:timeline1}
\end{figure}

We want to time-share the FPGA fabric by round-robin execution of
multiple realtime vision pipelines. Since every input frame needs to
be processed by every pipeline, initially we take \TframeP/{camera} to
be the basic scheduling quantum \Tround/ for one round of round-robin
execution. Each pipeline \PP/{i} is assigned a timeslice
\TslicePI/{P}{i}. During a pipeline's timeslice, the partitions
needed by the pipeline are configured first, and then one camera frame is fully processed. 

To present the same input frame to each
pipeline during its timeslice, the input frame from camera needs to be
double-buffered 1. to synchronize module execution after a partition reconfiguration with the start of every frame i.e. no frame skipping, and 2. for every pipeline to process every frame. \textit{We double-buffer input frame from camera into DRAM since the amount of data to buffer, which ranges from few KBs to MBs, may exceed the amount of available on-chip RAM (BRAM) on the FPGA we use (2.4MB)}.
During each timeslice, the runtime framework
drives the active pipeline with a pixel stream from DRAM at the
maximum rate the pipeline can handle or up to the DRAM
bandwidth. The output of the pipeline is also double-buffered into DRAM
so the runtime framework can produce an evenly timed output stream to display.
The multiple output video streams can be merged for display by a function
(e.g., XOR) or rendered simultaneously as split-screen.

\mypar/{Performance Model} If the total time to configure a
pipeline \PP/{i} is \TconfigPI/{P}{i},

\begin{center}
	\TslicePI/{P}{i} $=$ \TconfigPI/{P}{i} $+$ \TfillrawPI/{P}{i} $+$
	\TframerawPI/{P}{i}
\end{center}

\noindent 
Note in the above, \TfillNo/ and \TframeNo/ are for when the pipelines are operating against DRAM.  For a valid realtime schedule,

\begin{center}
	$\sum_{\text{all}~\sf{P}_i}$ \TslicePI/{P}{i} $\leq$ \Tround/ = \TframeP/{camera} 
\end{center}

\noindent

Figure~\ref{fig:timeline1} illustrates an execution timeline when two three-stage pipelines are time-shared as described above. This straightforward approach is not sufficient for achieving useful frame rates given the reconfiguration speed on today's FPGAs. The next section presents additional techniques needed to achieve usable performance.

%% file: 4_techniques.tex
\section{Reconfiguration Time Optimizations}
\label{sec:techniques}

As noted in the introduction, the time to reconfigure a RP is a few to
10s of milliseconds (Figure~\ref{fig:raw_overhead}), and multiple RPs
can only be configured one after the other. Therefore, the time
to reconfigure the RPs for a pipeline is often comparable with the
processing time. The time to reconfigure the RPs is also significant
relative to \TframeP/{camera}. If we use time-sharing as described in previous section, the time to configure a pipeline alone will
exceed \TframeP/{camera} in most non-trivial scenarios.  This section
introduces techniques to hide, amortize or eliminate the
reconfiguration time when possible.

\begin{figure}[t]
	\centering
	\includegraphics[width=0.5\textwidth]{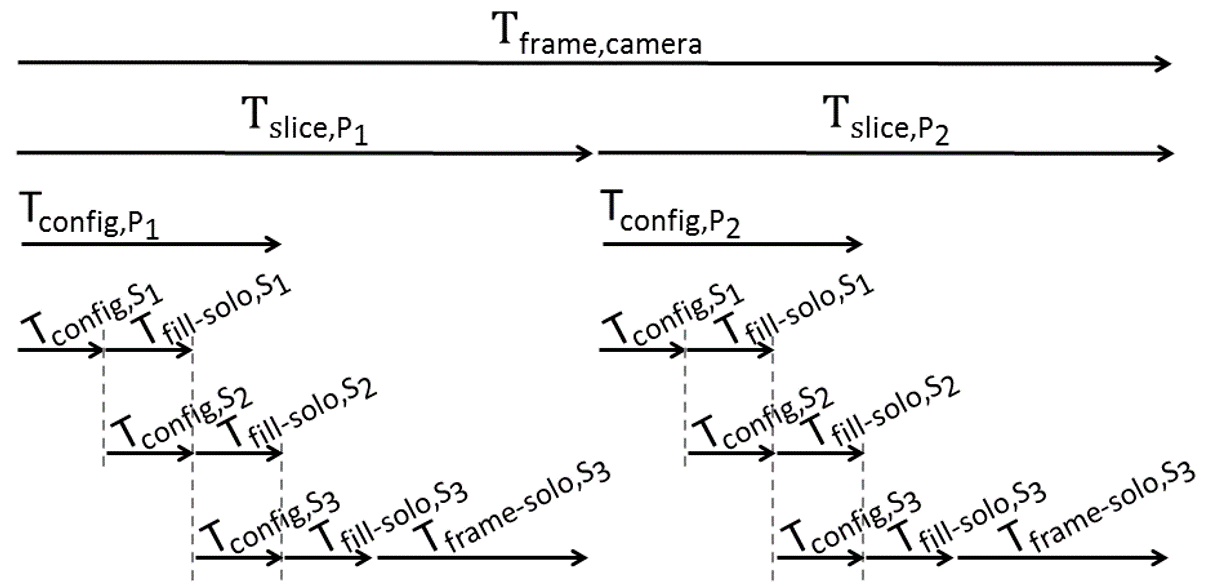}
	\caption{Time-sharing by two three-stage pipelines.  A stage of a pipeline
		starts processing as early as possible, that is, when the stage is configured AND its upstream stage is producing output.}
	\label{fig:timeline2}
\end{figure}

\subsection{Overlapping Reconfiguration and Processing}
\label{ssec:staggered}

In the last section, we waited until all of RPs of a pipeline have been
configured before starting processing. With all of the stages ready,
streaming processing can progress synchronously throughout the
pipeline. However, given that reconfiguration time of a partition is
significant relative to processing time, we are motivated to
overlap processing and reconfiguration by (1) reconfiguring RPs in order
from first to last; and (2) streaming input into the earlier
stages as soon as they are ready. Figure~\ref{fig:timeline2}
illustrates the execution timeline for the same two pipelines used in
Figure~\ref{fig:timeline1} but now starting a stage as soon as
possible, in other words, when the stage is configured and its upstream stage is producing output.

In this staggered-start execution, it is possible for an upstream
stage to start producing output before its downstream stage is
ready. Thus, it becomes necessary to
introduce buffering as a part of the streaming connection
abstraction between a downstream stage being reconfigured and its upstream stage to support staggered start. The
buffering capacity must be sufficient to capture all of the output of an
upstream stage until the downstream stage is ready. Data is buffered into DRAM since the amount of data may exceed BRAM capacity. \textit{Hence, to buffer and delay the data stream until the downstream stage is ready, we need to use a decoupling DMA engine between each downstream stage being reconfigured and its upstream stage.} In the worst
case, we have found it necessary to support the option for streaming
connections to be physically realized as a circular-buffer in
DRAM. This need for DRAM streaming connections motivates a more generalized streaming
interconnect infrastructure to connect RPs.

With staggered start, \TsliceP/{P} of pipeline \PP/{} is upper
bounded by

\begin{center}
	\TconfigP/{P} $+$ \TfillrawP/{P} $+$ \TframerawP/{P}
\end{center}

\noindent 
In the case when all the stages have comparable processing time
\TsliceP/{P} is lower bounded by

\begin{center}
	\TconfigP/{P} $+$ \TfillrawPI/{S}{\text{last}} $+$ \TframerawPI/{S}{\text{last}}
\end{center}

\noindent 
where \TfillrawPI/{S}{\text{last}} and
\TframerawPI/{S}{\text{last}} are \TfillrawNo/ and \TframerawNo/ of
the last pipeline stage only.  When some stages have much longer
configuration or processing time \TsliceP/{P} is more tightly lower
bounded by

MAX over all stages ${\sf{S}}_i$:

\begin{center}
	$\Big(\sum_{{\sf{S}}_j<{\sf{S}}_i}$ \TconfigPI/{S}{j}$\Big)$ + \TfillrawPI/{S}{i} + \TframerawPI/{S}{i}
\end{center}

\subsection{Amortization and Downsampling}

In the basic scheme presented in previous section, a round of round-robin
execution is completed for each quantum
\Tround/=\TframeP/{camera}. This is not necessary.  We can increase
\Tround/ to be a multiple $g\times$\TframeP/{camera}.  In this case,
we would double-buffer $g$ frames at a time from camera into DRAM.  During each
pipeline's timeslice, the runtime framework drives the active pipeline with
$g$ consecutive active frames from the DRAM double-buffer. Thus the
cost of reconfiguration is amortized over a longer processing time.
This option can be used with or without staggered start. In both
cases, \TslicePI/{P}{i} of pipeline \PP/{i} is now upper bounded by

\begin{center}
	\TconfigPI/{P}{i} $+$ \TfillrawPI/{P}{i} $+$ $g \times$ \TframerawPI/{P}{i}
\end{center}

\noindent 
For a valid realtime schedule,

\begin{center}
	$\sum_{\text{all}~\sf{P}_i}$ \TslicePI/{P}{i} $\leq$ \Tround/ = $g \times$ \TframeP/{camera} 
\end{center}

\noindent
The runtime framework can still produce a smooth video output when $g>1$
because the output is also double-buffered. Beside the added storage
cost, a major downside of increasing $g$ is the very
large increase in end-to-end latency through the runtime framework (which now
includes the time to buffer multiple frames of the input and output).

Also note, increasing $g$ improves scheduling by amortizing the
reconfiguration cost.  Therefore, it cannot help when the sum of the
pipeline's processing time already exceeds \TframeP/{camera}. In this
case, the only option is to downsample the video stream from camera into the pipeline. If the runtime framework
selectively only passes every $s$ frames of the camera input to the
pipelines, the pipeline timeslices only need to fit within the new
scheduling quantum of

\begin{center}
	\Tround/= $g \times s \times$ \TframeP/{camera}
\end{center}

\begin{figure}[t]
	\centering
	\includegraphics[width=0.5\textwidth]{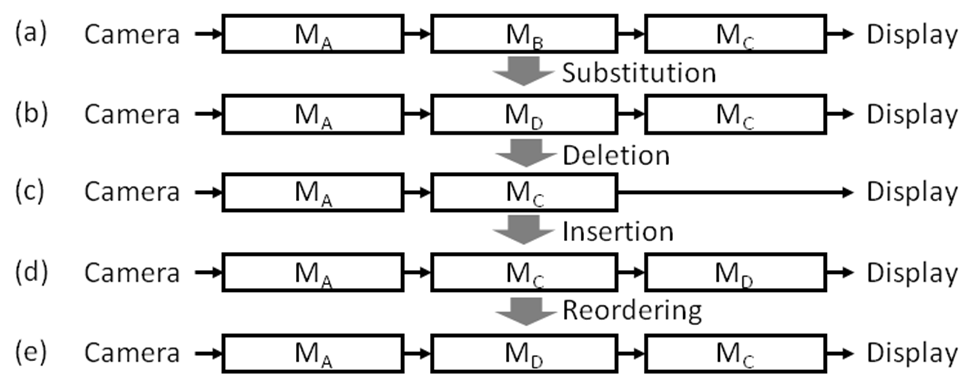}
	\caption{An illustrative example set of transitions between pipelines
		where reconfiguration of RPs can be avoided by retaining and reusing
		already configured stages.}  
	\label{fig:overview_tm}
\end{figure}

\subsection{Configurable Streaming Interconnect}
\label{ssec:interconnect}

In vision processing, even when pipelines have different
functionalities, they may share common stages (see for example
Figure~\ref{fig:pipe_eval}). The high cost of reconfiguring an RP can
be avoided when an already configured processing stage can be retained
and reused across pipelines.

Figure~\ref{fig:overview_tm} identifies different ways for multiple
pipelines to reuse common stage configurations. The simplest scenario
is when switching from pipeline (a) to pipeline (b) where the two
pipelines have the same topology and differ only in one stage. To
switch from (a) to (b) (and vice versa), only the middle RP has to be
reconfigured.  When switching from (b) to (c), no RP reconfiguration
is needed if there is a way to skip over the \MM/{D} stage of (b).
Furthermore, by retaining \MM/{D} even though it is not used by (c),
a switch from (c) back to (b) can also be done without
RP reconfiguration. On the other hand, also with \MM/{D} in place, a
switch from (c) to (d) does not require reconfiguration but
requires a different streaming connectivity than going to (b). In
fact, one can switch between (b), (c), (d) arbitrarily without
RP reconfigurations but as combinations of deletion, insertion, or
reordering by changing the connectivity between already
configured RPs.

To support these different scenario, we provide a flexibly configurable
streaming interconnect between RPs and other infrastructural elements.  A
configurable crossbar connects all elements in the system. This crossbar is not reconfigured by DPR bitstream but by control registers that can be quickly written by the controlling software between pipeline
reconfigurations to establish the desired static streaming topology for the
next timeslice. 
\textit{Configuring the interconnect by software is 3-orders of magnitude faster than reconfiguring an RP.} 
The statically
decided streaming connectivity topology never has two streaming
sources going to the same destination so the crossbar can be simple and
efficient with no need for flow control nor buffering.  (To support forks and joins in the pipeline, some RPs
have multiple input or multiple output interfaces while others have
exactly one input and one output
interface.)  For
simple streaming connections, the upstream and downstream stages
are connected with a single-cycle buffered path.  When a DRAM streaming
connection is used to allow for buffering and stage decoupling
(Section~\ref{ssec:staggered}), the source and sink RPs' streaming
interfaces are redirected to/from the DMA engines of the infrastructure instead.

This flexible interconnect infrastructure turned out to be a critical
mechanism. The expense of this interconnect infrastructure is well
justified by the reconfiguration time savings. As we will see in the
evaluation section, given the currently high cost of reconfiguration,
practical time-sharing is only feasible if the number of reconfigured
partitions between pipeline switches is kept to a minimum.

%% file: 5_prototype.tex
\section{Prototype System}
\label{sec:solution}

%\begin{figure}[t]
%	\centering
%	\includegraphics[width=0.3\textwidth]{figures/pic_setup}
%	\caption{Picture of a Xilinx ZC706
%		development board and a Vita 2000 sensor.}  
%	\label{fig:real_system}
%\end{figure}

\begin{table}[t]
	\small
	\begin{center}
		\caption {Xilinx ZC706 board specification} 
		\label{tab:FPGA_area} 
		\resizebox{0.3\textwidth}{!}{
		\begin{tabular}{ll}
			\hline 
			FPGA  & Xilinx XCZ7045\\ 
			Hard CPU Cores & 2 x ARM A9\\
			LUT & 218,600 \\
			BRAM (36 Kb) & 545 \\
			DSP & 900 \\
			DRAM Bandwidth & 12.8 GB/s (Fabric only)\\
			\hline
		\end{tabular} 
		}
	\end{center}
	\end {table}

%\begin{table}[t]
%	\small
%	\begin{center}
%		\caption {Interconnect (IC) vs RP reconfiguration cost  } 
%		\label{tab:IC_cost} 
%		%	\resizebox{\linewidth}{!}{
%		\begin{tabular}{l|l}
%			\hline 
%			One IC link  & few $\mu$s \\ 
%			\hline 
%		    One RP  & $> 1.5ms$ \\ 
%			\hline
%		\end{tabular} 
%		%	}
%	\end{center}
%	\end {table}

%\begin{table*}[t]
%	\small
%	\begin{center}
%		\caption {Reconfiguration cost for fixed and flexible stage-to-partition mapping using pipeline transitions from Figure \ref{fig:overview_tm} } 
%		\label{tab:fixed_flexible} 
%		%	\resizebox{\linewidth}{!}{
%		\begin{tabular}{l|l|l|l|l}
%			\hline 
%			& Substitution & Deletion & Insertion & Reordering\\ 
%			\hline 
%			Fixed mapping  & 2 RP  & RP  & RP & RP \\ 
%			\hline
%			Flexible mapping (our approach)  & Deletion & Insertion & Reordering\\
%			
%			\hline
%		\end{tabular} 
%		%	}
%	\end{center}
%	\end {table*}
%		
	We have implemented a working prototype of the time-sharing runtime framework.
	%(Figure~\ref{fig:real_system}).  
	The prototype system is built on a
	Xilinx ZC706 development board with an Xilinx XCZ7045 Zynq SoC FPGA.
	(Table~\ref{tab:FPGA_area} gives the specifications of this board.)
	Camera input comes from a VITA 2000-sensor that supports up to 1920-by-1080
	resolution at 60~fps (1080p@60fps). The prototype's HDMI video output can
	drive standard monitors. This section gives an overview of the
	prototype.
	
	% and provide a characterization of the overhead cost of the
	%framework infrastructure in performance and resource.
	
	\subsection{System Overview}
	
	\begin{figure}[t]
		\centering
		\includegraphics[width=0.29\textwidth]{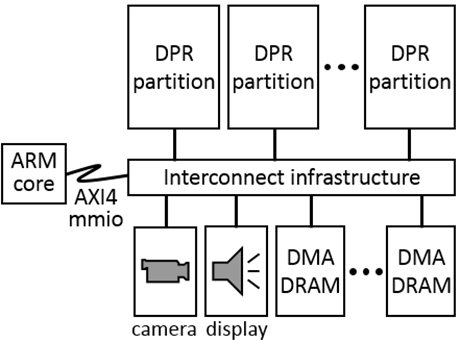}
		\caption{The high-level organization of the static runtime framework infrastructure.}
		\label{fig:high_level}
	\end{figure}
	
	\mypar/{Static Region Infrastructure} The organization of the
	runtime framework infrastructure implemented in the static region is shown in
	Figure \ref{fig:high_level}. The backbone of the runtime framework is the 
	configurable interconnect infrastructure discussed in
	Section~\ref{ssec:interconnect}. This interconnect infrastructure provides streaming connections
	between ten RPs for vision processing stages, the camera controller
	input, HDMI controller output, as well as five DMA engines for streaming
	to and from DRAM buffers. The interconnect is based on a custom
	crossbar implementation but the interface follows AXI4-Stream standard.
	Once configured, the interconnect infrastructure is capable of
	streaming frames at 1080p@60~fps between a fixed pair of
	source and sink RPs. Two RPs can also be connected by a
	DRAM streaming connection that incorporates a circular-buffer FIFO in DRAM.
	%In this case, the sink RP can back-pressure the source RP.
	Except for the camera and display controllers, the entire
	system--static region and reconfigurable partitions---are by default clocked
	at 200~MHz. The camera and display controllers are clocked at 148.5~MHz.
	
	\mypar/{Management Software} At runtime, a runtime manager running on
	the embedded ARM processor core manages the creation, execution and
	time-sharing of vision pipelines. The specification of each pipeline
	(such as number of stages, module running in each stage and
	connectivity between stages) is registered with the runtime manager. To
	switch execution to a new pipeline, the runtime manager assigns a stage
	to an RP if the RP already has the required module. The RPs for the
	remaining stages are reconfigured through the PCAP interface with bitstreams loaded from DRAM. Once the partitions are reconfigured, the runtime manager configures the modules, DMA engines, and interconnect to effect the required
	connectivity before starting pipeline execution. The
	built-in camera and display controllers are initialized once when the FPGA is first started.
	
	For time-sharing, the runtime manager will cycle through all of the
	registered pipelines once for every $g$ frames of video. The runtime
	manager will poll the active pipeline for completion before initiating
	a switch to the next pipeline. This runtime manager does not do
	scheduling or enforce maximum time quantum. If total time to cycle through
	all of the pipelines exceeds the time quantum of $g \times s
	\times $\TframeP/{clock}, the processing falls out of sync to produce
	glitching output.
	
	\mypar/{Vision Modules} We use Xilinx Vivado HLS to develop
	custom modules. We also make use of the HLS video library that offers
	a subset of HLS-synthesizable OpenCV functions. These HLS-based modules can be incorporated into our runtime framework since our interconnect supports
	AXI4-streaming interface.
	
	%We wait for a module to finish processing a frame by polling it before
	%starting the reconfiguration process which consists in 1. stopping
	%interactions between the partition and the rest of the system
	%2. sending a request to the PCAP to reconfigure the partitions with
	%requested module stored in external DRAM 3. configuring the
	%interconnect infrastructure for connecting the partition to the rest
	%of the system 4. starting the module that has been loaded into the
	%p[artition 5. resuming the interactions between the partition and the
	%rest of the system. When multiple partitions are reconfigured, we
	%iterate the reconfiguration process as many times as needed.
	
	%We use Vivado version 2015.2 for synthesis, placement \& routing and
	%bitstream generation of both the static system and module library. We
	%use Xilinx SDK version 2015.2 for running our system on a standalone
	%ARM CPU.
	
	%The framework uses DPR to support the dynamic creation, execution and
	%interleaving of multiple vision pipelines. A pipeline consists of one
	%or more stages mapped to partitions. Partitions are reconfigured with
	%desired modules and then connected to each other through the
	%interconnect infrastructure to form the requested pipeline. When
	%pipelines can not be all mapped simultaneously on the fabric (due to
	%lack of resources), we interleave their execution at a frame
	%granularity. The operation of the framework (partition
	%reconfiguration, partition connectivity and pipeline interleaving) is
	%managed from software running on an embedded processor core.
	
	\subsection{Runtime Framework Characteristics} 
	
	%%this table is in the last section
	\begin {table}[t]
	\begin{center}
		\caption {Logic resource used by static region and reconfigurable partitions on the Xilinx XCZ7045.}
		\label{tab:resource_system} 
		\resizebox{0.5\textwidth}{!}{
			\begin{tabular}{l|lll||l}
				& & \bf{Static} & & \bf{Reconfigurable}\\
				\hline
				\hline
				& Crossbar  & DMA engines & Misc & \\
				LUT & 4940 (2\%) & 10725 (5\%) & 30578 (14\%) & 122400 (56\%)\\ 
				BRAM (36 Kb) & 0 & 15 (3\%) & 23.5 (4\%) & 360 (66\%)\\
				DSP & 0 & 0 & 0 & 300 (33\%) \\
			\end{tabular} 
		}
	\end{center}
	\end {table}
	
	\mypar/{Logic Resource Utilization} Table~\ref{tab:resource_system}
	breaks down the fabric resource utilization between the static region and
	reconfigurable partitions. The infrastructure logic requires non-trivial
	resources. The interconnect crossbar is only a small fraction of the
	total infrastructure. On the other hand, the DMA engines to stream
	data through DRAM is quite expensive. 
	
	On a large FPGA like the Xilinx XCZ7045, ample resources remain to be divided as ten independent
	reconfigurable RPs. We aimed for a total fabric utilization of
	roughly 70\% to ease the placement and routing process.

	\mypar/{DRAM Bandwidth} The DRAM bandwidth on Xilinx ZC706 development board is
	12.8~GB/s. This bandwidth is shared by all of the DRAM streaming 
	connections through AXI HP ports. To support 1080p@60fps, each DRAM streaming connection requires a total of 497 MB/sec of  memory bandwidth (read and write). DRAM streaming connections include the double-buffers for the camera input and display output, and the
	decoupling buffers needed to support staggered start
	(Section~\ref{ssec:staggered}). 
	
\begin{figure}[t]
		\centering
		\includegraphics[width=0.4\textwidth]{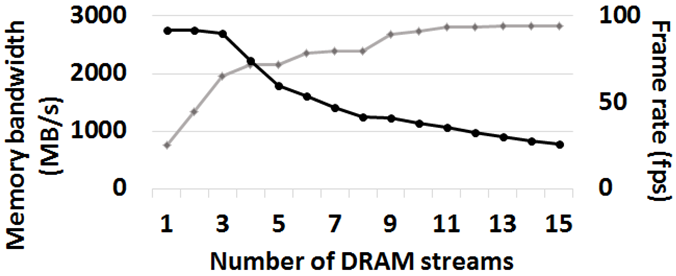}
		\caption{Measured DRAM bandwidth utilized vs the number of DRAM streaming connections on the ZC706 board. Up to five DRAM streaming connections can concurrently sustain 1080p@60fps. }
		\label{fig:mem_bandwidth}
	\end{figure}

	We created a microbenchmark to measure the total DRAM bandwidth actually
	utilized for increasing number of active thru-DRAM streaming
	connections. The results are shown in
	Figure~\ref{fig:mem_bandwidth}. On the Xilinx ZC706 development board,
	only up to 5 concurrent thru-DRAM streaming connections can be
	supported for 1080p@60fps. Since two thru-DRAM streaming
	connections are taken up by camera and display for double-buffering,
	we are left with only three usable thru-DRAM streaming connections for
	decoupling the staggered start of RPs
	(Section~\ref{sec:techniques}). This restricts the
	applicability of the staggered start optimization in the prototype.
	
	%the Xilinx ZC706 development board.

	%%streams (read and write), up to five system components (reconfigurable
	%%partitions and IO controllers) get sufficient memory bandwidth to
	%%achieve up to full HD rate. In our system, subtracting two DRAM
	%%streams (one for writing camera output to DRAM and providing data for
	%%display, and one allocated for the partition before the one being
	%%reconfigured), up to three partitions can be reconfigured
	%%sequentially.
	%%
	%%Three DRAM streams allow up to three single-stage pipelines to run
	%%concurrently by decoupling partitions. While one pipeline is paused
	%%during a reconfiguration, the output of a running pipeline can be
	%%displayed instead. The constraint on maximum number of DRAM streams is
	%%platform-dependent. Using a board with more bandwidth, the number of
	%%DMA streams could easily scale up.
	%%
	%%
	%%
	%%specification
	%%The frameowrk manaThe multi-function vision system has 10
	%%reconfigurable partitions and an interconnection infrastructure with
	%%15 endpoints connecting DMA engines and reconfigurable
	%%partitions \footnote {The interconnection infrastructure, DMA engines
	%%  and IO controllers (camera and display) are part of the static
	%%  region.}

%% file: 6b_results.tex
\section{Performance Evaluation}
\label{sec:results}

This section presents an application-level evaluation of the
time-sharing runtime framework prototype. This evaluation aims to show that useful realtime performance (30+~fps) can be achieved when time-sharing multiple streaming vision pipelines. We first quantify the opportunities of using DPR for repurposing and realtime time-sharing.
We then present our results when measuring the achieved performance of time-shared pipelines in frames-per-second (fps) under different operating conditions with the camera running at 720p@60~fps and 1080p@60~fps. 

%present the pipelines used in our case study and motivate the need for using a DPR design flow versus a static design flow. We then present our experimental setup and our results when measuring the achieved performance of time-shared pipelines in frames-per-second (fps) under different operating conditions with the camera running at 720p and 1080p resolution at 60~fps. 

\begin{figure*}[t]
	\centering \includegraphics[width=1\textwidth]{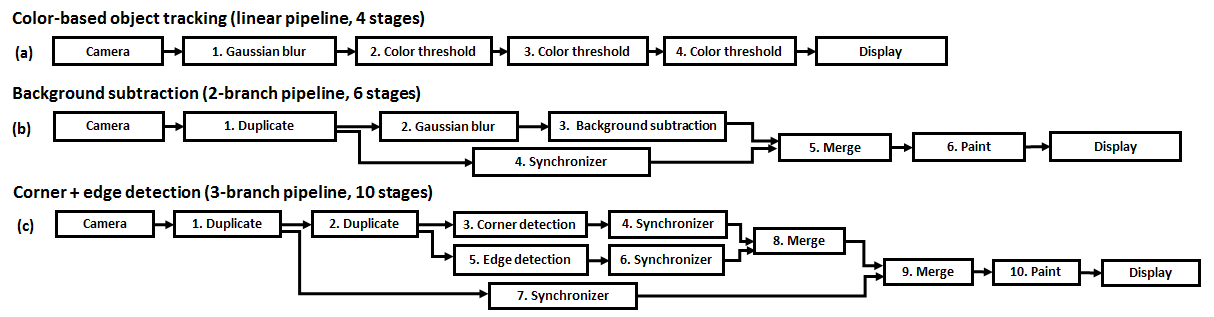}
	\caption{Logical view of three pipeline examples:
		(a)~color-based object tracking where objects of up to three different colors are tracked, (b)~background subtraction, (c)~corner and edge detection. %This view does not correspond to the physical mapping of stages to RPs.  %The stages highlighted in grey are not common to all pipelines. When time-sharing, some of them need to be reconfigured while others can be inserted or deleted with our interconnect infrastructure. 
	}
	\label{fig:pipe_eval}
\end{figure*}

\begin {table*}[!h]
\begin{center}
	\caption {Vision modules used in our evaluation.}
	\label{tab:modules} 
	\resizebox{0.9\textwidth}{!}{
		\begin{tabular}{l||l}
			Edge detection & computes a binary mask of vertical and horizontal edges using a Sobel filter~\cite{itseez2014theopencv}	\\		
			\hline		
			Color-based object tracking & tracks objects based on their color \\
			\hline
			 Template tracking & tracks a given template by computing sum-of-absolute differences and by thresholding\\		
			\hline
			Corner detection & computes a binary mask of corners using a Harris corner detector~\cite{itseez2014theopencv}
			\\
			\hline
			Blob detection &  detects blobs by using morphological operations and thresholding 
			\\
			\hline
			Gaussian blur & blurs an image by using a Gaussian filter
			\\
			\hline
			Background subtraction & removes frame background by thresholding 
			\\
		\end{tabular} 
	}
\end{center}
\end {table*} 

\begin {table*}[!h]
\begin{center}
	\caption {Logic resource used by seven 3-branch pipelines. Each pipeline is mapped individually in a static FPGA design (without runtime framework), and runs at 250~MHz.}
	\label{tab:area_ind} 
	\resizebox{1\textwidth}{!}{
		\begin{tabular}{l||l|l|l|l|l|l|l}
			& edge + corner & edge+template & blob+color &	edge+color & corner+template & background + corner  & background + edge	\\		
			\hline	
			\hline		
			LUT & 13147	& 13098 & 14142 & 13601 & 14085 & 14797 &	13810 \\
			
			FF & 12455 & 11635 & 11423 & 11234 & 12146 & 13222 & 12711 \\		
			
			BRAM (36 Kb) & 5 &5 & 3.5 & 3.5 & 5 & 3.5 & 3.5
			\\		
		\end{tabular} 
	}
\end{center}
\end {table*}

\subsection{Opportunities for Using DPR}

\mypar/{Interactive Realtime Vision Applications} The dynamic adaptation requirement of interactive realtime vision systems leads to a large number of potential pipelines to execute at runtime. The realtime vision system we built (introduced in \ref{sec:intro}) requires the flexible, dynamic creation and execution of a large number of pipelines. These pipelines can have a variable topology and number of stages. Figure~\ref{fig:pipe_eval} shows the logical view of three pipeline examples that have different number of stages and topology. (A non-linear pipeline topology is standard in vision and allows to overlay different masks, computed on each branch, on the original camera frame.) Each pipeline branch can execute one or a combination of vision modules listed in Table \ref{tab:modules}, leading to hundreds of potential different pipelines. 
%In the examples from Figure~\ref{fig:pipe_eval}, a non-linear pipeline forks into two or three branches after the second stage, where each branch processes one of the duplicated streams. On each of the upper branches, the pipeline computes a mask to be merged later with either a second mask or the original frame (on the bottom branch) to produce an output frame with the highlighted mask(s). The bottom branch delays the pixels of the original image (synchronizer stage) to account for \TfillrawNo/ so they can be later merged with their corresponding pixels in the computed mask(s). Similarly, when the pipeline has three branches, the computed mask on each branch is delayed before merging.

%\begin{itemize}
%	\item edge detection: computes a binary mask of vertical and horizontal edges using a Sobel filter~\cite{itseez2014theopencv}
%	\item color-based object tracking: tracks objects based on their color
%	\item template tracking: tracks a given template by computing sum-of-absolute differences and by thresholding
%	\item corner detection: computes a binary mask of corners using a Harris corner detector~\cite{itseez2014theopencv}
%	\item blob detection: detects blobs by using morphological operations and thresholding 
%	\item Gaussian blur: blurs an image by using a Gaussian filter
%	\item background subtraction: removes frame background by thresholding 
%\end{itemize}
%Each pipeline overlays one mask computed on the top branch (or two merged masks computed in the first two branches) on the original frame.
\mypar/{Static Design Limitation} Traditionally, one way to implement such a system is to map all pipelines simultaneously on the FPGA.
%Statically mapping all pipelines requires a large device, and, more problematically, the knowledge of all possible pipelines to run at runtime ahead of time. 
Table \ref{tab:area_ind} presents the logic resources used by seven of the most resource-expensive 3-branch pipelines, when each pipeline is mapped individually and directly on the FPGA (without DPR). These numbers give an idea of the potential cost of mapping a large number of parallel pipelines statically on an FPGA. If all possible linear and non-linear pipelines were to be mapped statically and simultaneously, they would not fit on the FPGA. Mapping those pipelines individually and directly to the FPGA results in the best possible performance. We expect performance to degrade with an increasing number of parallel pipelines to map statically. When mapped individually on the FPGA, each of these pipelines can make timing at 250~MHz. 

\mypar/{DPR Performance Results} 
DPR presents a viable alternative to overcome the inflexibility and the resource limitation of a static FPGA design.
%for designing an interactive realtime vision system. DPR allows the flexible, dynamic creation and execution of a large number of pipelines by overcoming the cost limitation of a static FPGA design flow. 
When using DPR, we expect the performance to degrade compared to static design due to the RP I/O placement port constraint that can add wire delay. For repurposing and realtime time-sharing, the performance needs to be sufficient for correct realtime pipeline operation for 720p@60fps and 1080p@60fps input video. Also, there should be enough performance slack to interleave pipelines at the time scale of a camera frame.

To assert the performance of a DPR system, we use the system described in section \ref{sec:solution}. The ten RPs are differently sized to support repurposing and time-sharing. The four largest RPs (bitstream size of 1.1MB) are reconfigured when repurposing while the six smallest RPs (bitstream size of 300KB) are used for time-sharing. We generate partial bitstreams for the seven modules such that all modules can be hosted in any RPs. We are able to generate
partial bitstreams at 200~MHz when our runtime framework is imposed. Despite the expected performance degradation, pipelines can operate correctly for 720p@60fps and 1080p@60fps input video (when run by themselves in the runtime framework without time-sharing). An operating speed of 200~MHz also allows to time-share pipelines at the time scale of a camera frame. 

%In the rest of this section, we evaluate the performance  and show that we can achieve useful performance when time-sharing. 

\subsection{Reconfiguration Overhead}

Before presenting the performance of time-shared pipelines, we need to simplify the set of pipelines that we use for the time-sharing evaluation. To do so, we perform a first set of experiments to assert that the cost of switching from one pipeline to another is dominated by the cost of RP reconfiguration i.e. the cost of configuring the interconnect, the DMA engines, the modules, and the cost of starting the pipeline are negligible. In these experiments, the pipelines occupy up to ten stages and have up to three branches. 

%To verify that the cost for reconfiguring the interconnect is negligible, 
We generate randomly tens of pipeline pairs (with different topology, different stages, different number of stages etc), and measure the time to interleave two pipelines in a pair. 
We find that the sole overhead that matters is the time spent in RP reconfiguration. RP reconfiguration dominates the cost of a pipeline switch by three orders of magnitude. (The cost of reconfiguring the interconnect for topology change, and other configuration and startup cost, are within the range of 50s to 100s of microseconds depending on the number of RPs and interconnect links to reconfigure.) Switching between pipelines with different topology does not impact time-shared performance. If switching from one pipeline to another does not change the state of any RP, the switch is almost free. (This is the case, for instance, when the set of stages used by one pipeline is a subset of the stages used by the other pipeline.)
%In our case,  

\subsection{Performance of Time-Shared Pipelines}

For this evaluation, we only consider linear pipelines since pipeline topology does not impact performance as established previously. For these experiments, pipelines occupy up to six stages, and two interleaved pipelines differ by one, two, three, four, five and six RPs. (We only reconfigure the six smallest RPs while the four largest RPs remain unchanged. The time spent in reconfiguring RPs is proportional to the number of RPs to reconfigure since the six RPs have the same size.) When time-sharing, we execute pipelines (1)~two at a time, and (2)~three at a time.  The runtime framework logic produces a simultaneous split-screen video output of the time-shared pipelines.  

\begin{figure}
	\centering \includegraphics[width=0.50\textwidth]{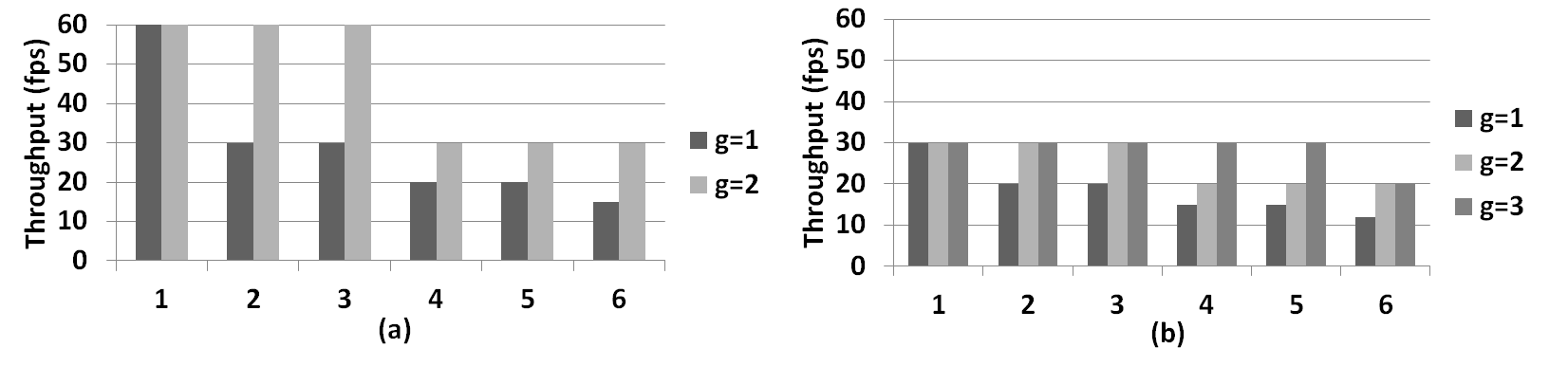}
	\caption{The frames-per-second (fps) for each time-shared pipeline for 720p@60~fps input video when we execute (a) two pipelines (b) three pipelines at a time. We reconfigure between one to six RPs per pipeline switch. Each time-shared pipeline processes $g$ consecutive frames before reconfiguration.}
	\label{fig:perf_720}
\end{figure}

Figure~\ref{fig:perf_720} first summarizes the achieved performance in frames-per-second when the runtime framework is driven with a 720p@60fps video stream, and when we execute (a) two pipelines at a time (b) three pipelines at a time by time-sharing. In Figure~\ref{fig:perf_720}.a and Figure~\ref{fig:perf_720}.b, there are six sets of bars corresponding to cases where we reconfigure between one to six RPs to switch from one pipeline to another. For
each case, bars for different $g$ are shown. Figure~\ref{fig:perf_720}.a shows the
processing time required by two pipelines can fit into the
$g=1$ (every frame is processed) scheduling quantum of 16.7~milliseconds (=$\frac{1}{60}$~second) when one RP is reconfigured per pipeline transition. 
Factoring in reconfiguration time for more than one RP, the time-shared
execution of the two pipelines reconfiguration can only keep up when the input is
downsampled by $s=2$, i.e., each pipeline runs at 30~fps (video output at 30~fps is
still visually smooth), by $s=3$, i.e., each pipeline runs at 20~fps, or by $s=4$, i.e., each pipeline runs at 15~fps. Running the runtime framework at $g=2$ (two consecutive frames are processed)
can restore the frame rate of each pipeline to 60~fps with up to three RP reconfigurations. Factoring in reconfiguration time for more than three RPs, the time-shared
execution of the two pipelines reconfiguration can only keep up when the input is
downsampled by $s=2$, i.e., each pipeline runs at 30~fps. 

When time-sharing by
three pipelines (figure~\ref{fig:perf_720}.b), the interleaved execution only keeps up for $g=1$ when the input
is downsampled by $s=2$ (one RP reconfiguration), i.e., 30~fps, or by $s\ge3$ (more than one RP reconfiguration), i.e., $\le30~fps$. Increasing $g$ to 3 in this
case allows $s$ to be reduced to 2 (except for the last case when six RPs are reconfigured). 

\begin{figure}
	\centering \includegraphics[width=0.50\textwidth]{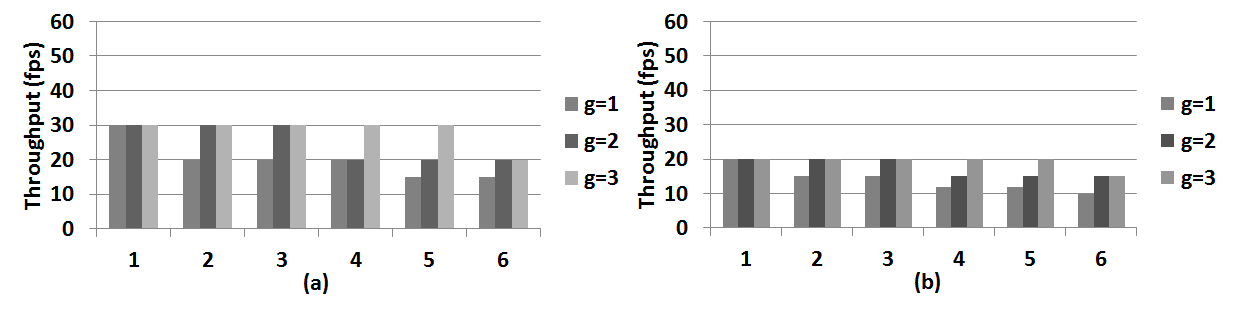}
	\caption{The frames-per-second (fps) for each time-shared pipeline for 1080p@60~fps input video when we execute (a) two pipelines (b) three pipelines at a time. We reconfigure between one to six RPs per pipeline switch. Each time-shared pipeline processes $g$ consecutive frames before reconfiguration.}
	\label{fig:perf_1080}
\end{figure}

Figure~\ref{fig:perf_1080} similarly summarizes the achieved
performance measured in frames-per-second when the runtime framework is driven with a 1080p@60fps
video stream, and when we execute (a) two pipelines at a time (b) three pipelines at a time by time-sharing. For 1080p processing, the higher processing time required
by two pipelines, without considering reconfiguration time, already
would not have fit into the $g=1$ scheduling quantum of
16.7~milliseconds. In this case, increasing $g$ cannot improve
scheduling slack. Thus, time-shared execution of two pipelines (figure~\ref{fig:perf_1080}.a)
requires downsampling by $s\ge2$, i.e., $\le30~fps$. Time-shared execution of
three pipelines (figure~\ref{fig:perf_1080}.b) would require further downsampling to $s\ge3$, i.e.,
$\le20~fps$.

%% file: 7_discussion.tex
\section{Discussion and Conclusion}
\label{sec:discussion}

This paper has discussed and demonstrated the feasability of using DPR for time-sharing despite the restrictive reconfiguration time.  
We developed four techniques that are part of an essential set to overcome high reconfiguration time when possible. We demonstrated through a working runtime framework the practical feasibility of
timing-sharing by vision pipelines at useful frame
rates---30~fps for 1080p and 60~fps for 720p on the Xilinx ZC706 board.

In this paper, we examined the opportunity of using DPR for time-sharing in vision systems, and showed that time-sharing is a promising approach in cases where commonality can be exploited (stage sharing across pipelines), and where performance can be traded for flexibility and cost (in terms of device and power cost). If maximum performance is required for a fixed set of functionalities known ahead of time, mapping all pipelines simultaneously and statically is preferred.

In addition to our proposed techniques, orthogonal approaches have been discussed in the literature and could be used in our system to further reduce reconfiguration time, as in \cite{6780588}. A disadvantage of using PCAP is that the CPU blocks until a partition reconfiguration is done. The time wasted in blocking is not an important limitation in our case since the CPU is not a compute stage and can only use this time to reconfigure the interconnect (tens of microseconds).
%In our case, if the CPU was not blocked, the interconnect could be reconfigured (in this work, the CPU is not a compute stage).  the time to reconfigure one interconnect link is in the order of microseconds. 

In our work, the scheduling
and mapping of stages to RPs rely on manual decision. A scheduler that can manage resource and time with higher resolution is part of future work. %are not only far from optimal but also rely on manual decision making quite extensively.  

%However, if the set of non-tightly performance-constrained applications is rapidly changing and/or growing, time-sharing is a more cost-effective approach (in terms of device and power cost). 
%Another time-sharing use case is the use of multiple cameras where each camera stream feeds multiple different streaming pipelines. 

%Still, applications that require time-sharing at time quanta less than 10s of milliseconds is
%beyond the reach of today's FPGAs.
Faster and concurrent DPR support in the future would improve
effectiveness of time-sharing. DRAM bandwidth is another limitation when we need to buffer streams in DRAM. %as ways to 
%increase flexibility and performance. 
These improvements are
critical to enable time-sharing applications at time quanta less than 10s of milliseconds.

%future-generation computing-oriented FPGAs as
%full-fledged computing devices.

%Our working runtime framework realization  
%*scale up limited by bandwidth and configuration ports (talk about resolution) i.e. talk about number of ports and projection if more were available
%*sure you could put everything on big FPGA but benefit  is being to do more in smaller devices when not everything is needed simultaneously. If more functionalities are needed, need to scale more and more
%*if more filters were reconfigured, we would miss a time quantum => show the point where you would miss the thing
%*power impact => not so much
%* acknowledge other techniques for reduction of reconfiguration time but explain that these techniques can be used in conjunction with what we have (combitgen) 
%* sure custom PCAP controller  could be done, but this is part of future work => we can project and say yes, that would improve
%* sure, for now, the call to PCAP is blocking but processor is not doing useful computation and reconfiguring the IC is actually very fast.

%% file: 8_acknowledgments.tex
\section{Acknowledgments}
\label{sec:Acknowledgments}

This work was supported in part by the CONIX Research Center, one of six centers in JUMP, a Semiconductor Research Corporation (SRC) program sponsored by DARPA.

%% file: main.bbl
\begin{thebibliography}{10}

\bibitem{1303106}
M.~Ullmann, M.~Huebner, B.~Grimm, and J.~Becker, ``An fpga run-time system for
  dynamical on-demand reconfiguration,'' in {\em 18th International Parallel
  and Distributed Processing Symposium, 2004. Proceedings.}, pp.~135--, April
  2004.

\bibitem{xilinx}
Xilinx, ``{{Vivado Design Suite User Guide: Partial Reconfiguration
  (UG909)}},'' 2016.

\bibitem{Majer:2007:ESM:1265130.1265134}
M.~Majer, J.~Teich, A.~Ahmadinia, and C.~Bobda, ``{{The {{Erlangen}} Slot
  Machine: A Dynamically Reconfigurable {{FPGA}}-based Computer}},'' {\em J.
  VLSI Signal Process. Syst.}, vol.~47, pp.~15--31, Apr. 2007.

\bibitem{autovision}
C.~Claus, W.~Stechele, and A.~Herkersdorf, ``Autovision – a run-time
  reconfigurable mpsoc architecture for future driver assistance systems
  (autovision – eine zur laufzeit rekonfigurierbare mpsoc architektur für
  zukünftige fahrerassistenzsysteme),'' vol.~49, pp.~181--, 05 2007.

\bibitem{Koch:2011:FHP:1950413.1950427}
D.~Koch and J.~Torresen, ``{FPGASort: A High Performance Sorting Architecture
  Exploiting Run-time Reconfiguration on Fpgas for Large Problem Sorting},'' in
  {\em Proceedings of the 19th ACM/SIGDA International Symposium on Field
  Programmable Gate Arrays}, FPGA '11, (New York, NY, USA), pp.~45--54, ACM,
  2011.

\bibitem{6128547}
D.~Goehringer, L.~Meder, M.~Hubner, and J.~Becker, ``{{Adaptive Multi-client
  Network-on-Chip Memory}},'' in {\em 2011 International Conference on
  Reconfigurable Computing and FPGAs}, pp.~7--12, Nov 2011.

\bibitem{6339136}
C.~H. Hoo and A.~Kumar, ``{{An area-efficient partially reconfigurable crossbar
  switch with low reconfiguration delay}},'' in {\em 22nd International
  Conference on Field Programmable Logic and Applications (FPL)}, pp.~400--406,
  Aug 2012.

\bibitem{Arram:2015:RRA:2684746.2689066}
J.~Arram, W.~Luk, and P.~Jiang, ``{{Ramethy: Reconfigurable Acceleration of
  Bisulfite Sequence Alignment}},'' in {\em Proceedings of the 2015 ACM/SIGDA
  International Symposium on Field-Programmable Gate Arrays}, FPGA '15, (New
  York, NY, USA), pp.~250--259, ACM, 2015.

\bibitem{Niu:2015:EOC:2684746.2689076}
X.~Niu, W.~Luk, and Y.~Wang, ``{{EURECA: On-Chip Configuration Generation for
  Effective Dynamic Data Access}},'' in {\em Proceedings of the 2015 ACM/SIGDA
  International Symposium on Field-Programmable Gate Arrays}, FPGA '15, (New
  York, NY, USA), pp.~74--83, ACM, 2015.

\bibitem{6861604}
S.~Byma, J.~G. Steffan, H.~Bannazadeh, A.~L. Garcia, and P.~Chow, ``{{FPGAs in
  the Cloud: Booting Virtualized Hardware Accelerators with OpenStack}},'' in
  {\em 2014 IEEE 22nd Annual International Symposium on Field-Programmable
  Custom Computing Machines}, pp.~109--116, May 2014.

\bibitem{4630015}
C.~Claus, W.~Stechele, M.~Kovatsch, J.~Angermeier, and J.~Teich, ``A comparison
  of embedded reconfigurable video-processing architectures,'' in {\em 2008
  International Conference on Field Programmable Logic and Applications},
  pp.~587--590, Sept 2008.

\bibitem{itseez2014theopencv}
Itseez, {\em The OpenCV Reference Manual}, 2.4.9.0~ed., April 2014.

\bibitem{6780588}
K.~Vipin and S.~A. Fahmy, ``Zycap: Efficient partial reconfiguration management
  on the xilinx zynq,'' {\em IEEE Embedded Systems Letters}, vol.~6,
  pp.~41--44, Sept 2014.

\end{thebibliography}
